\documentclass[8pt]{article}

\usepackage[english]{babel}
\usepackage{geometry}
\usepackage{amsmath}
\usepackage{graphicx}
\usepackage{soul}
\usepackage[colorlinks=true, allcolors=blue]{hyperref}

\title{Artificial Theory of Mind and Self-Guided Social Organisation}
\author{Michael S. Harr\'e, Jaime Ruiz-Serra, Catherine Drysdale}

\begin{document}
\date{}
\maketitle

%\begin{abstract}
%Your abstract.
%\end{abstract}

%\section{Introduction}

One of the challenges artificial intelligence (AI) faces is how a collection of agents coordinate their behaviour to achieve goals that are not reachable by any single agent~\cite{wolpert1999introduction,levin2023bioelectric}. In a recent article by Ozmen et al~\cite{ozmen2023six} this was framed as one of six grand challenges: That AI needs to respect human cognitive processes at the human-AI interaction frontier. We suggest that this extends to the AI-AI frontier and that it should also {\it reflect} human psychology, as it is the only successful framework we have from which to build out. In this extended abstract we first make the case for collective intelligence in a general setting, drawing on recent work from single neuron complexity in neural networks and ant network adaptability in ant colonies. From there we introduce how species relate to one another in an ecological network via niche {\it selection}, niche {\it choice}, and niche {\it conformity} with the aim of forming an analogy with human social network development as new agents join together and coordinate. From there we show how our social structures are influenced by our neuro-physiology, our psychology, and our language. This emphasises how individual people within a social network influence the structure and performance of that network in  complex tasks, and that cognitive faculties such as Theory of Mind play a central role. We finish by discussing the current state of the art in AI and where there is potential for further development of a {\it socially embodied collective artificial intelligence} that is capable of guiding its own social structures.

Biological agents are composed of sub-units such as organs, cells, and molecular networks~\cite{levin2019computational}, in which, for example, individual neurons display computational intelligence, or ``competencies", that can scale freely~\cite{levin2022technological}. Beniaguev et al.\cite{beniaguev2021single} demonstrated that a single cortical neuron requires five to eight layers in a deep neural network to approximate its input-output mapping, and so we may conceptualise a neuron as an artificial agent with complex computational abilities that adapts its connections to other agents. Similarly, ant colonies have conserved social structures and division of labor across species separated by over 100 million years of evolution~\cite{kay2024ant}, with leadership roles in individual ants enhancing collective performance~\cite{richardson2021leadership}. In response to pathogenic threats, ants adapt by reorganizing their social behaviors~\cite{stockmaier2021infectious} and modifying their nests to slow disease spread~\cite{leckie2024architectural}. While neural networks exhibit ``solid brain" connectivity (agent-to-agent connectivity is rigid in time) and ant colonies demonstrate ``liquid brain" dynamics~\cite{sole2019liquid} (agent-to-agent interactions are fluid in time), both systems continuously integrate information and adjust their connectivity for the collective benefit.

There are however differences in how freely new agents connect in these two model systems. The adult brain has limited capacity for neurogenesis so that recruitment, placement, and integration of new cells plays a relatively small role in adult mammals. But in liquid brains new agents can freely join collectives and contribute to the collective outcomes of joint actions. The process by which this occurs in ecological network is detailed in~\cite{muller2020power}, where the interaction between a single plant, for example, and its ecological environment results in individualised niches via three main processes: niche choice, niche conformance, and niche construction. Niche choice occurs when an individual selects environmental conditions that align with its phenotype, while niche conformance involves adjusting its phenotype to suit the environment. Niche construction is the modification of the environment to meet individual needs, which may also impact other species~\cite{clark2020niche}. These processes affect both individual, improving the match between the individual and its ecological network, thereby improving their overall fitness. Effective niche realisation relies on communication between encoders-senders and receivers-decoders using so-called {\it info-chemicals} as the signalling mechanism. In sum, interactions between individual plants and animals, communicated over information carrying channels, shape network relations at all levels in ecological networks.

This is also the case in the context of human social networks, where, as we argue later, Theory of Mind (ToM) has a key role to play. In his review of Darwin's {\it strange inversion of reasoning}, Dennet~\cite{dennett2009darwin} opens with the following: 
\begin{quote}
    Darwin's theory of evolution by natural selection unifies the world of physics with the world of meaning and purpose by proposing a deeply counterintuitive ``inversion of reasoning'' (according to a 19th century critic): ``to make a perfect and beautiful machine, it is not requisite to know how to make it''~\cite{beverley1867darwinian}.
\end{quote}
Here we suggest that this was inverted again by the evolution of our very human capacity to be {\it consciously} aware of ourselves and our social and cultural constructs, and to {\it purposefully} manipulate them to our collective benefit. Take this observation about manipulating inter-cellular communication by Watson and Levin~\cite{watson2023collective}: 
\begin{quote}
    This framework [of collective cellular intelligence] makes a strong prediction: if intercellular signalling (not genes) is the cognitive medium of a morphogenetic individual, it should be possible to exploit the tools of behavioural and neuro-science and learn to read, interpret and re-write its information content in a way that allows predictive control over its behaviour (in this case, growth and form) without genetic changes. 
\end{quote}
A counter question then arises: How do humans {\it read, interpret, and re-write our inter-personal information content} in order to exert predictive control over a social group? Similar to how a scientist \textit{external} to a cell collective manipulates inter-cellular signaling to control collective outcomes, individuals \textit{internal} to a human collective can influence inter-personal behaviors to (re)shape social structures in order to alter group outcomes. In both cases, an agent with a goal-directed psychology manipulates inter-agent relationships---whether inter-cellular or inter-personal---to control higher-level outcomes, be it at the organism or societal scale. And so, as new people join a social group and just as new species form niches, rather than integrating a new person based on mechanisms such as random connectivity or rich-get-richer processes, a dynamic integration occurs whereby the networked agents and the prospective agent selectively adjust relationships to accommodate (or reject) the newcomer. This requires complex psychological abilities in order to do it at the time scale of the phenotype, rather than at the evolutionary time-scale, which we address later in this work. It is achieved via a highly evolved set of  cognitive tools that humans achieve this, as we discuss next.

ToM is the ability to mentally represent the internal states of another person~\cite{frith2010social} and a surprising connection exists between the development of ToM and language. The {\it representational} view suggests that both children~\cite{farrar2002early} and adults use specific grammatical structures to represent complex events and reason from them~\cite{de2021role}. These {\it complement structures} are an example of language-as-cognitive tool that has special utility in representing the mental states of other people, for example they allow for the expression of mistakes, lies, or false beliefs. Complement structures are strong predictors of children's false belief understanding (a canonical test of ToM) in longitudinal studies, and training in complement syntax has been shown to improve children's false belief reasoning. It is interesting to note that in developmental learning there is a stronger effect from language to ToM ability than vice versa~\cite{milligan2007language}.

Language and ToM also strongly interact with our causal reasoning in social settings. Lombard and Gärdenfors~\cite{lombard2023causal} put forward three hypotheses: (1) ToM is a crucial component of causal cognition; (2) The more sophisticated causal cognition becomes, the more it relies on ToM; and (3) The evolution of causal cognition increasingly depends on mental representations of hidden variables. The aspect we wish to focus on here is the causal modelling. At the individual level, recent work by Momennejad~\cite{momennejad2022collective} reviewed neuro-imaging evidence for people neurologically encoding their social network topologies and the sharing of these encodings across a social group. This was also demonstrated in the work of Lau et al.~\cite{lau2018discovering} showing that people are able to integrate information about how agents relate to one another in addition to how they relate to oneself in order to infer social group structures. This allows agents to have a collective causal understanding of how individuals socially interact with one another. Finally, there is also evidence for improved collective intelligence when individuals with higher competencies in ToM are present in a social setting~\cite{woolley2010evidence}. From this we conclude that language, a shared collective understanding of social causal relationships, and our ToM are part of a highly integrated cognitive toolbox that we have developed to understand how we collectively fit together to improve our coordination towards collective goals. We emphasise that the collective goals are also psychological constructs of individuals, and in pursuit of these goals people take advantage of their ability to read, interpret, and even re-write the social connections between themselves and other agents in order to achieve a greater collective outcome than any single agent could achieve by themselves.

Where does artificial intelligence sit with respect to these cognitive tools? Inverse reinforcement learning (IRL) has been suggested as a model for ToM in AI~\cite{jara2019theory,ruiz2023inverse} where an agent infers the preferences, e.g. rewards as a function of state, of another agent based solely on its behaviour. Jara-Ettinger~\cite{jara2019theory} discusses some key limitations of IRL but misses the social network within which the agents are situated and the causal cognitive model of social interactions. Large Language Models (LLMs) have also been a very popular test-bed for ToM in AI~\cite{strachan2024testing}, however it has been shown that while LLMs can replicate causal reasoning in some scenarios, they also have highly unpredictable failure modes~\cite{kiciman2023causal} and there are very few studies on {\it causal social cognition}. There is also no evidence to date that the complete suite of cognitive skills mentioned above have been demonstrated by an LLM. In particular, there are no LLMs that are socially embodied in a communication network whereby an individual agent articulates a collective goal, manipulates the social connections between other agents in order to optimise the network structure in order to achieve that goal, and to do so based on a shared representation of each agent's causal model as even early humans could do~\cite{haidle2010working}. However, there has been some promising advances made in formally progressing the state of the art, for example in~\cite{watson2011global}, where Watson et al. show that
\begin{quote}
    ... when self-interested agents can modify how they are affected by other agents (e.g., when they can influence which other agents they interact with), then, in adapting these inter-agent relationships to maximize their own utility, they will necessarily alter them in a manner homologous with Hebbian learning. Multi-agent systems with adaptable relationships will thereby exhibit the same system-level behaviors as neural networks under Hebbian learning.
\end{quote}
There are many other avenues that are being explored in the artificial context that we cannot cover in this extended abstract, but in our talk at GSO-2025 we will cover some of the major themes that have been contributing to this research and where their strengths and weaknesses lie. In broad strokes our presentation will take a leaf from Shevlin and Halina and urge caution in applying rich psychological terms in AI~\cite{shevlin2019apply}.
%This is a cognitive technology that we argue is likely necessary but not sufficient for people to learn to interpret, share and re-write interpersonal communications. 

%Human cognitive structures,
%Human planning and foresight,
%We cognitively encode our social networks,
%Theory of Mind: Its recent recasting in artificial intelligence and human psychology,
%Language and Theory of Mind: Co-constructors of social meaning

\bibliographystyle{unsrt}
\bibliography{sample}

\end{document}